\documentclass{PoS}
\usepackage{xcolor}
\usepackage{graphicx}
\usepackage{booktabs}
\usepackage{amsmath}
\usepackage{array}
\usepackage{multirow}

\makeatletter

\newcommand*{\@rowstyle}{}

\newcommand*{\rowstyle}[1]{
  \gdef\@rowstyle{#1}%
  \@rowstyle\ignorespaces%
}

\newcolumntype{=}{
  >{\gdef\@rowstyle{}}%
}

\newcolumntype{+}{
  >{\@rowstyle}%
}

\makeatother

\title{2+1 Flavor Domain Wall Fermion QCD Lattices:  Ensemble Production and (some) Properties}

\ShortTitle{2+1 Flavor DWF Lattices}

\author{\speaker{Robert D. Mawhinney}\thanks{This work was done in
collaboration with members of the RBC and UKQCD Collaborations,
with major contributions from Chulwoo Jung, David Murphy and Jiqun
Tu.  Substantial support for the implementation in QUDA of some of
the algorithms discussed here was provided by Kate Clark of Nvidia.
Time on the Summit computer at ORNL was provided through DOE INCITE
award PHY131 in 2019.  MSPCG algorithm
development and algorithm tuning on the small volume was done on
Mira at the ALCF.  This research used resources of the Oak Ridge
Leadership Computing Facility at the Oak Ridge National Laboratory,
which is supported by the Office of Science of the U.S. Department
of Energy under Contract No. DE-AC05-00OR22725.
This research used resources of the Argonne
Leadership Computing Facility, which is a DOE Office of Science
User Facility supported under Contract DE-AC02-06CH11357.  The
author is supported in part by U.S.\ DOE grant \#DE-SC0011941.}\\
	Columbia University \\
        The RBC and UKQCD Collaborations\\
        E-mail: \email{rdm10@columbia.edu}}

\abstract{The RBC and UKQCD Collaborations continue to produce 2+1
flavor domain wall fermion ensembles, currently focusing on an
ensemble with a $96^3 \times 192$ volume on SUMMIT at ORNL with
$1/a \approx 2.8$ GeV, and smaller ensembles at stronger couplings.  The
$1/a \approx 2.8$ GeV ensemble uses the Exact One Flavor Algorithm for
the strange quark, along with the Multisplitting Preconditioned
Conjugate Gradient for solving the Dirac equation.  We report on
our progress and experience to date with the evolution of this
ensemble.}

\FullConference{37th International Symposium on Lattice Field Theory - Lattice2019\\
		16-22 June 2019\\
		Wuhan, China}

\begin{document}

\section{Introduction}

The RBC and UKQCD Collaborations have been generating 2+1 flavor Domain Wall
Fermion (DWF) and Mobius Domain Wall Fermion (MDWF) ensembles since 2005.
A complete list of these ensembles, with the Iwasaki gauge action, is given in Table 1.
These have been used for measurements for many different QCD 
observables, including kaon decays, measurements of LECs for chiral perturbation
theory, the hadron vacuum polarization and hadronic light-by-light scattering
for $g-2$ of the muon, and the $K_L - K_S$ mass difference.  (We note that
another set of 2+1 flavor ensembles, with the Iwasaki + Dislocation Supressing
Determinant Ratio (DSDR) gauge action is also available.  The presence
of the DSDR term allows for simulations at larger lattice spacings, while still
allowing the residual mass to be kept small for accessible values of the fifth
dimension, $L_s$, of (M)DWF.)

For the last 5 years, 2 ensembles with spatial volumes greater than
(5.4 fm)$^3$ and essentially physical pion and kaon masses have
been available \cite{DWF-PHYS}.  These are shown in the first two red rows in Table
1.  Combining these ensembles at two lattice
spacings with a third Iwasaki plus DSDR ensemble, and requiring a
common continuum limit for observables, has shown that the ${\cal
O}(a^2)$ scaling errors are modest and that, within our statistics,
only ${\cal O}(a^2)$ correction terms are needed.

However, for any observables dependent on the valence charm quark
mass, the coarsest Iwasaki action ensemble ($1/a = 1.730$ GeV) and
the Iwasaki+DSDR ensemble ($1/a = 1.35$ GeV) are too coarse to be
of use, even for simulations with $m_c$ lighter than its physical
value, where an eventual extrapolation to the physical $m_c$ value
is done.  For access to charm quark physics, a smaller lattice
spacing is needed to augment the $1/a = 2.359$ GeV ensemble.  Such
an ensemble was produced, with $1/a = 2.774$ GeV, $m_\pi = 234$ MeV
and a (3.5 fm)$^3$ spatial volume \cite{DWF2774}.  In spite of the smaller lattice
spacing, reasonable topological tunneling was observed for this
ensemble with current algorithms.  Given this success, we have begun the
production of an ensemble with this lattice spacing, physical pion mass and a
larger volume.  This ensemble's properties (expected
from out global chiral perturbation theory fits to all of our other
ensembles) are shown in the last red row of Table 1.  The ongoing
production of lattices in this ensemble is the subject of this
report.

\begin{table}[hbt]
\begin{center}
\begin{tabular}{=l +c +c +c +c +c +c +c +c} \toprule
	Action &$1/a$ & Lattice          &  $m_l$ & $m_s$ & $m_{\rm res}$ & $m_\pi$ & $m_K$ & Size \\ \cmidrule(rl){4-6}
	(F+G)  &(GeV) & volume           &  \multicolumn{3}{c}{(in lattice units)}   & (MeV)     & (MeV)    & (fm) \\ \midrule
DWF+I &1.785(5) & $24^3 \! \times \! 64 \! \times \! 16$ & 0.03 &0.04  & 0.00308 & 693 &  & 2.6   \\
DWF+I &1.785(5) & $24^3 \! \times \! 64 \! \times \! 16$ & 0.02 &0.04  & 0.00308 & 576 &  & 2.6   \\
DWF+I &1.785(5) & $24^3 \! \times \! 64 \! \times \! 16$ & 0.01 &0.04  & 0.00308 & 432 & 626 &  2.6   \\
DWF+I &1.785(5) & $24^3 \! \times \! 64 \! \times \! 16$ & 0.005&0.04 & 0.00308 & 340 & 593 & 2.6   \\
\rowstyle{\color{red}}
MDWF+I &1.730(4) & $48^3 \! \times \! 96 \! \times \! 24$ & 0.00078&0.0362  & 0.000614 & 139 & 499 &5.5  \\ \\

%
DWF+I &2.383(9) & $32^3 \! \times \! 64 \! \times \! 16$ & 0.008 & 0.03 & 0.000664 & 412 & 615 & 2.6  \\
DWF+I &2.383(9) & $32^3 \! \times \! 64 \! \times \! 16$ & 0.006 & 0.03 & 0.000664 & 360 & 596 & 2.6  \\
DWF+I &2.383(9) & $32^3 \! \times \! 64 \! \times \! 16$ & 0.004 & 0.03 & 0.000664 & 302 & 579 & 2.6  \\
\rowstyle{\color{red}}
MDWF+I &2.359(7) & $64^3 \! \times \! 128 \! \times \! 12$ & 0.000678&0.02661  & 0.000314 & 139 & 508 & 5.4  \\ \\
MDWF+I & 2.774(10) & $48^3 \! \times \! 96 \! \times \! 12$ & 0.002144 & 0.02144  & 0.000229 & 234 & 516 & 3.5  \\
\rowstyle{\color{red}}
MDWF+I & 2.774(10) & $96^3 \! \times \! 192 \! \times \! 12$ & 0.000541 & 0.0213  & 0.000229 & 135 & 495 & 6.9 \\ \\
DWF+I  & 3.15(2) & $32^3 \! \times \! 64 \! \times \! 12$ & 0.0047 & 0.0186  & 0.000631 & 371 & 558 & 2.0  \\ \bottomrule
\end{tabular}
\end{center}
	\caption{A summary of the 2+1 flavor (M)DWF ensembles generated by the
	RBC and UKQCD Collaborations.  All ensembles use the Iwasaki gauge action.
	The three ensembles highlighted in red have essentially physical pion
	and kaon masses and spatial volumes greater than (5.4 fm)$^3$.  This report
	focuses on the ongoing production of the $96^3 \times 192 \times 12$ ensemble
	with $1/a = 2.774(10)$ GeV.}
\end{table}

\section{Using The Exact One Flavor Algorithm}

Using standard pseudofermion methods to represent the determinant of a single
quark flavor has been possible for many years, with the advent of the
Rational Hybrid Monte Carlo (RHMC) algorithm.  Generically, $\det(D)$, which represents
a single flavor must be changed to $\det(D^\dagger D)$, which represents
two flavors, in order to have an operator with a positive definite spectrum
which is suitable for representation by non-Grassmanian (bosonic) pseudofermion fields.
To get back to the correct determinant factor for a single flavor,
one needs to use pseudofermions to represent $\det([D^\dagger D]^{1/2})$.
The RHMC algorithm uses a rational function representation to accurately
represent the square root function over the spectral range of the eigenvalues
of $D^\dagger D$.

While the RHMC has been very successful, the single quark flavor
(strange) that it implements in modern (M)DWF simulations has become
computationally costly, in spite of the large value of the strange
quark mass in comparison to the degenerate light quark masses.
There are two primary reasons for this.  First, while the required
Dirac equation solves for the RHMC can be performed with a multi-shift
conjugate gradient algorithm, this solver is not restartable, so
it is difficult to run in lower precision.  Since modern CPUs and
GPUs are memory bandwidth limited and have high-performance, reduced
precision floating point units, there is a substantial performance
penalty for running the conjugate gradient (CG) partly or largely in
double precision.  Secondly, for (M)DWF, adding Hasenbusch
preconditioning masses, {\em i.e.}\ 
$\det([D^\dagger(m_1)D(m_1)]^{1/2})$/$\det([D^\dagger(m_2)D(m_2)]^{1/2})$ is
expensive, since the 1/2 power of the Dirac operators appearing in
the denominator must be implemented as two, 1/4 power operators to
preserve reversibility of the algorithm.  From the light quark
sector, it is well known that Hasenbusch preconditioning is important
in balancing the size of the fermionic force the integrator must
handle with the expense of calculating that force.  The lack of
Hasenbusch preconditioners for the RHMC part of the algorithm
increases the expense of this part of the simulation.

Fortunately, an alternative to the RHMC, the Exact One Flavor
Algorithm (EOFA) has been proposed in \cite{EOFA1} and
shows that
\begin{equation}
	\det \left[ \frac{D_{DWF}(m_1)}{D_{DWF}(m_2)}\right] 
	= \frac{1}{\det{\cal M}_L} \frac{1}{\det{\cal M}_R}
\end{equation}
with ${\cal M}_L$ and ${\cal M}_R$ Hermitian and postive definite.
We have tested and implemented this algorithm as detailed
in \cite{EOFA-RBC} and have found that the Dirac equation
solution needed for the EOFA system can be recast into
an appropriate solve for a (M)DWF system.  This allows us
to reuse the existing (M)DWF high performance solver code for
the EOFA system.  (For our G-parity simulations with MDWF,
the EOFA has been extremely beneficial, since here the
1/2 power of the light quark determinant is needed.
By adding in Hasenbusch preconditioning masses, being
able to solve in single precision with defect correction
restarts and retuning the parameters for the HMC, we
achieved a $4.5\times$ wall clock speed up over our earlier
RHMC implementation.)

For the evolution of our current $96^3 \times 192 \times 12$ ensemble
with $1/a = 2.774(10)$ GeV on Summit, we have implemented the EOFA
algorithm on GPUs.  In particular, the pseudofermion heat-bath for
the EOFA algorithm does require a rational approximation to a
fractional power of the EOFA operator.  (We stress that this is
only in the initial heat-bath part of a molecular dynamics trajectory.)
Implementation of this heat bath in QUDA was done by David Murphy.
The QUDA Mobius Dirac operator code had to be generalized to
implement the generic EOFA linear system and this was done by
Jiqun Tu.  Kate Clark of Nvidia provided invaluable advice
and support in these implementations.

\section{Algorithm Tuning}

In preparation for the generation of our $96^3 \times 192 \times
12$ ensemble on Summit, we thermalized a $32^3 \times 64 \times 12$
ensemble on Mira, with the same input quark masses and couplings.
This smaller ensemble has too small a volume, (2.3 fm)$^3$, for
sensible physics results, but for tuning algorithm parameters and
comparing the results with the RHMC and EOFA algorithms, it is
expected to be a reliable testing ground.

Tuning of the HMC algorithm involves making choices for the Hasenbusch
masses, choosing the conjugate gradient stopping condition for each
required Dirac equation solve, choosing the accuracy for the
conjugate gradient solves needed to calculate the initial and
final values for the Hamiltonian and choosing the step size to
achieve the desired acceptance (usally around 90\%).  Most of these
choices cannot be made until one has reasonably thermalized the
ensemble, since, for example, the fermion force due to each
Hasenbusch mass ratio generically depends on the thermalization.
In order to proceed, we start with a reasonable set of guesses for
these parameters and let the ensemble begin to thermalize.  Given
the much lower calculational cost for the smaller volume, it is
much easier to begin these tunings on this system.

We have found that the choice of Hasenbusch masses can be guided
by monitoring the L-infinity norm of $F\cdot dt$ over the entire lattice
for each step in the molecular dynamics.  Here $F$ is the fermionic 
force from a particular Hasenbusch determinant ratio and we plot
histograms of these L-infinity norms for some number of trajectories.
(Detailed examples of these tunings are given in \cite{EOFA-RBC}.)
Adding more Hasenbusch masses decreases the overall force, but
there is a cost to each additional mass.  We also find that the
acceptance decreases when there are occasional large forces and
that these fluctuations in forces are more pronounced for
the Hasenbusch ratios for the lightest quarks.

Our input light quark mass should be 0.000541 which adds to the expected
residual mass of 0.000229 to give a total light quark mass of 0.000770.
(The residual mass value comes from the $1/a = 2.744$ GeV ensemble
with a 234 MeV pion.  This is our best estimate for this and since
the residual mass has mild dependence on the input light quark mass,
we expect this to be reliable.)  From the small volume ensemble running,
after thermalizing for many hundreds of trajectories and experimenting
with different Hasenbusch choices, we gathered 56 trajectories of data
with intermediate Hasenbusch masses of 0.0056, 0.028, 0.1, 0.28,
0.56.  The forces from the determinant ratios containing the
masses (0.000541,0.0056) and (0.028, 0.1) showed occasional large
fermionic forces almost twice as large as the largest values
from other determinant ratios.

At this point, our Summit code was ready to run and the machine was
available, so we replicated our $32^3 \times 64 \times 12$ thermalized
lattice 3 times in each space-time direction (81 copies overall)
and began evolution of a $96^3 \times 192 \times 12$ ensemble on
Summit.  Clearly, substantial thermalization time is needed for
the large ensemble to eliminate the replications, but we can
continue our algorithm tuning during this time and the tunings
should be more reliable than tuning on an ensemble which
has not been equilibrated at all.

Figure 1 shows a histogram of the L-infinity norm of $F \cdot dt$
for 10 trajectories on the $96^3 \times 192 \times 12$ ensemble
after about 300 trajectories of thermalization on the large
volume after it was made by replicating the small volume.
An additional intermediate Hasenbusch mass has been added
and one can see that the occurence of large forces is fairly
similar for all of the Hasenbusch mass ratios.  One sees
that the lightest mass produces somewhat more large forces
than the others.  The figure also shows the forces due to
an intermediate Hasenbusch mass used in the EOFA algorithm.
One clearly sees that to keep the EOFA forces in agreement
with the light quark forces, an intermediate mass of 0.163
has been added.  Without the addition of this mass, the single-flavor
strange quark forces would be larger than those from any of the
individual light-quark Hasenbusch ratios.

\begin{figure}
\begin{center}
\includegraphics[scale=0.75]{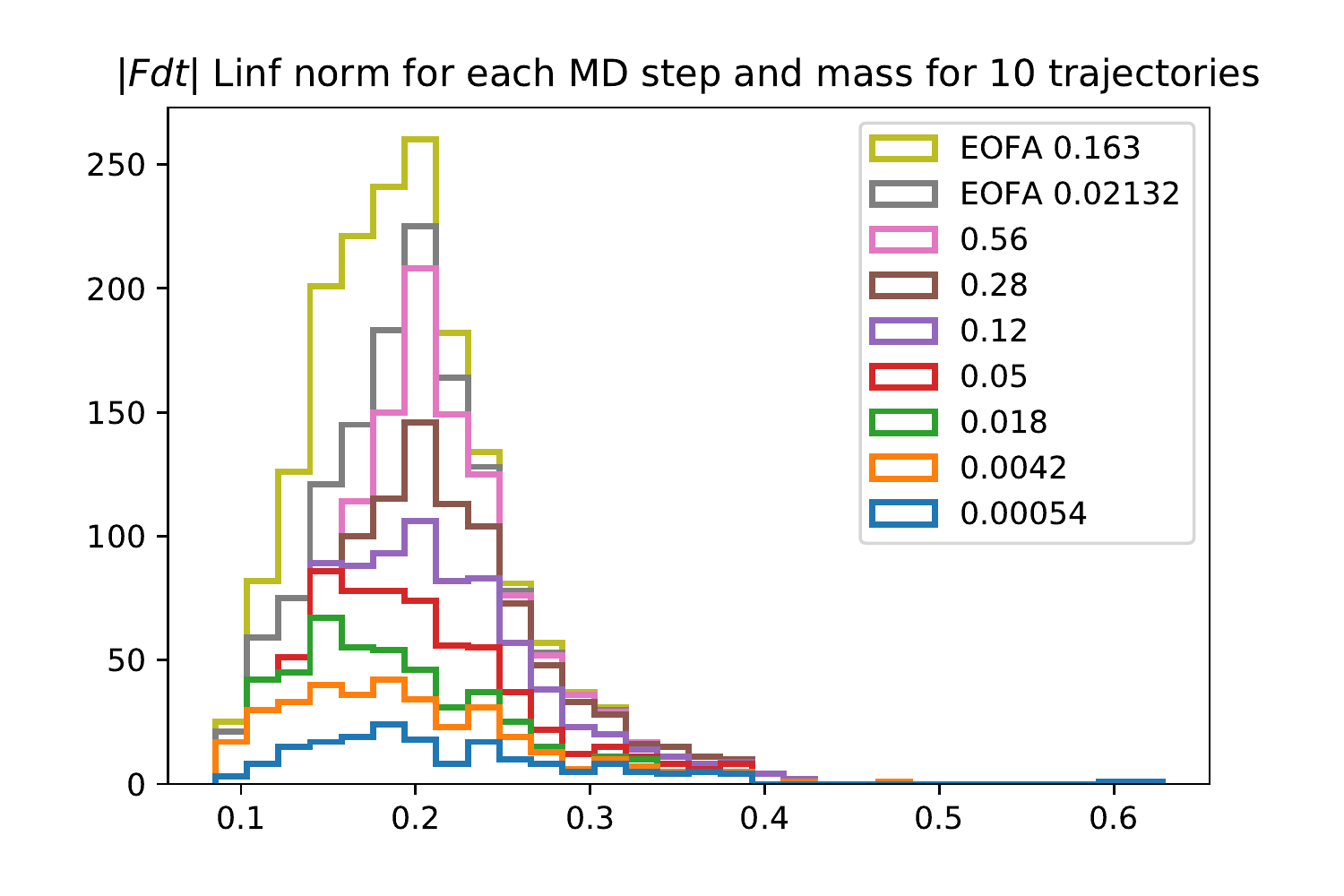}
	\caption{A histogram of the L-infinity norm of $F \cdot dt$ for the
	$96^3 \times 192 \times 12$ ensemble for 10 trajectories around trajectory 300.
	The number labeling a color is the quark mass appearing in the numerator
	of the Hasenbusch mass ratio. Unless explicitly labeled EOFA, the forces
	are for the light quarks.}
\end{center}
\end{figure}

Another test done on the small volume ensemble was to compare the speed
with the RHMC and EOFA algorithms.  With one choice of Hasenbusch masses,
and all other parameters held the same, just replacing the RHMC with the
EOFA decreased the time per trajectory by 30\%.  The overall speed up is
expected to be much larger, as is evident in Figure 1, since the EOFA
forces are now in line with the light-quark forces.  We have not tried
running any RHMC comparison tests for the $96^3 \times 192 \times 12$ ensemble,
given the costs of such a test and the need to keep acceptances the same
to fully understand the wall clock time differences.

\section{The Multisplitting Preconditioned Conjugate Gradient (MSPCG)} 

The Summit computer, with 6 Nvidia V100 GPUs per node, has extensive
on-node floating point power (42 TFlops, double precision), but
only a 100 GB/s Mellanox link providing network bandwidth between
nodes.  This is very little bandwidth for strong scaling the solvers
used for DWF QCD to, say, 1024 nodes, since the local volumes implied
by a 1024 node job generally require of order one byte of off-node
bandwidth per sustained Flop.  To increase local (on-node) floating point
utilization in the (M)DWF conjugate gradient, we have developed the
Multisplitting Preconditioned Conjugate Gradient (MSPCG) \cite{MSPCG,MSPCG-JT}. 
The Multisplitting algorithm \cite{MS} provides general criteria for
detailing how a linear equation solve can be split into submatrix pieces,
with each solved separately, and then an update step is done, which spans
the submatricies, to redefine the next iteration of the problem.

While investigating this method for the precondioned normal (M)DWF
operator
\begin{equation}
	D^\dagger_{PC} D_{PC} = [M_5 - M^4_{e} M^{-1}_5 M^4_{eo}]^\dagger
	                        [M_5 - M^4_{e} M^{-1}_5 M^4_{oe}]^\dagger
\end{equation}
which connects fourth nearest-neighbor sites on the lattice (see \cite{MSPCG,MSPCG-JT})
we originally failed to get convergence.  Our initial approach split
the underlying four-dimensional Wilson Dirac operator ($M_4$) into submatrices
localized on each GPU,
{\em i.e.}\  we tried to split the parts of $D^\dagger_{PC} D_{PC}$ into
submatrices rather than splitting $D^\dagger_{PC} D_{PC}$ into
submatrices.  Changing so that we were splitting $D^\dagger_{PC} D_{PC}$ into
submatrices made the Multisplitting algorithm converge, but it
converged more slowly than the original CG.  However, we were able to use
the local submatrix decomposition as a preconditioner to the CG
and found that the CG iteration count fell by a factor of 3 for
physical light quark masses \cite{MSPCG}.  (Using the multisplitting algorithm
as a preconditioner for the CG, when the submatrix decomposition is
the natural one given by the underlying lattice geometry, makes
our implementation the same as using the additive Schwarz
algorithm as a preconditioner.)

For our ensemble generation on Summit, the speed-up possible
with the MSPCG depends on the performance that can be achieved
for a strictly local linear equation solve on each V100.
If these local solves were infinitely fast, {\em i.e.}\ the
preconditioner took no time, our evolution would speed up
by almost a factor of 3.  The preconditioner allows us to
trade local floating point power for network bandwidth.

Jiqun Tu has implemented the preconditioner in the QUDA CG, with
support and consultation provided by Kate Clark of Nvidia.
Despite their large floating point speed, the memory
bandwidth of the V100s limits their performance and hence
the speed of the preconditioner.  Tu's implementation of
the preconditioner has also utilized the tensor cores on the
V100.  These are very efficient for dense matrix multiply
and the fifth-dimensional part of (M)DWF does contain a 
substantial amount of dense matrix arithmetic. By fusing kernels
to keep data in the device memory, Tu's implementation
is able to see a performance boost through the use
of the tensor cores.

Table 2 summarizes the performance of the MSPCG on the
target $96^3 \times 192 \times 12$ ensemble for various
numbers of nodes and CG steps for the preconditioner.
One important item to note is that, for the 1024 node
case, the standard CG runs at 2.93 TFlops/node, which is
about 2 times faster than the performance when Summit first
came on line.  This reflects optimizations and improvements 
in the Summit software stack.  For this same 1024 node
case, MPSCG gives a speed-up of 1.22, which is an important,
but modest gain.  If the outer CG was running at its
original slower speed, the gain from the MSPCG would
be much larger.

\begin{table}[h]
\scriptsize
\centering
\begin{tabular}{c|c|c|c|c|c|c|c|c}
\hline
\hline
nodes & local volume & solver & inner iter. & (outer) iter. & r.u. & performance/node & time & speed up \\
\hline
\multirow{ 5}{*}{$256$} & \multirow{ 5}{*}{$16\cdot24\cdot12\cdot24$}  & CG    & $-$ & $42133$ & $471$ & $4.66$                & \colorbox{red!30}{$486.3$} & \multirow{ 5}{*}{$1.10$x} \\
&   & MSPCG & $05$ & $16903$ & $195$ &$1.56(01)$/$5.45(35)$/$37.29(53)$  & $456.0$ \\
&   & MSPCG & $06$ & $14860$ & $173$ &$1.56(01)$/$5.51(31)$/$37.60(58)$  & \colorbox{blue!30}{$442.6$} \\
&   & MSPCG & $07$ & $13787$ & $161$ &$1.56(01)$/$5.48(28)$/$37.49(60)$  & $460.2$ \\
&   & MSPCG & $08$ & $12922$ & $151$ &$1.56(01)$/$5.44(26)$/$37.55(63)$  & $469.5$ \\
\hline
\multirow{ 5}{*}{$512$} & \multirow{ 5}{*}{$16\cdot12\cdot12\cdot24$}  & CG    & $-$ & $42427$ & $474$ & $3.85$                & \colorbox{red!30}{$296.6$} & \multirow{ 5}{*}{$1.13$x} \\
&   & MSPCG & $05$ & $17625$ & $203$ &$1.26(01)$/$4.54(37)$/$36.21(52)$  & $271.0$ \\
&   & MSPCG & $06$ & $15425$ & $179$ &$1.27(01)$/$4.55(33)$/$36.26(57)$  & \colorbox{blue!30}{$262.1$} \\
&   & MSPCG & $07$ & $14409$ & $168$ &$1.26(01)$/$4.57(30)$/$36.39(60)$  & $268.3$ \\
&   & MSPCG & $08$ & $13597$ & $159$ &$1.27(01)$/$4.53(28)$/$36.35(63)$  & $276.0$ \\
\hline
\multirow{ 5}{*}{$1024$} & \multirow{ 5}{*}{$16\cdot12\cdot12\cdot12$} & CG    & $-$ & $42482$ & $474$ & $2.93$                & \colorbox{red!30}{$195.2$} & \multirow{ 5}{*}{$1.22$x} \\
& & MSPCG & $05$ & $18250$ & $210$ &$1.00(01)$/$3.68(34)$/$34.62(45)$  & $183.3$ \\
& & MSPCG & $06$ & $15959$ & $185$ &$1.01(01)$/$3.68(35)$/$34.79(54)$  & \colorbox{blue!30}{$159.7$} \\
& & MSPCG & $07$ & $14985$ & $174$ &$1.01(01)$/$3.68(32)$/$35.06(58)$  & $163.6$ \\
& & MSPCG & $08$ & $14287$ & $167$ &$1.00(01)$/$3.69(29)$/$34.76(61)$  & $169.1$ \\
\hline
\hline
\end{tabular}
\caption{Strong scaling of the MSPCG on SUMMIT for solving the Dirac
	equation $D^\dagger_{PC} D_{PC} x=y$ to the accuracy of $10^{-12}$ on the
$96^3\times 192\times 12$, 2+1 flavor M\"obius domain wall fermion,
$1/a = 2.774$ GeV lattice with physical pion mass. Here $y$ is a
gaussian random source vector and the MSPCG is implemented within
	the QUDA software environment. The times are time-to-solution given
in units of seconds. The column titled "r.u." denotes the
	number of reliable updates performed.
In the "performance/node" column, the
performance is given in tera-flops per node, with the percentage
	of time spent in a particular part of the algorithm given
	in parentheses.  For the CG
solver the performance is given as the total performance, including
precise and sloppy dslash and linear algebra operations. For MSPCG
solves the performance is given in the format of precise/sloppy/preconditioned
dslash with their respective time percentages in parentheses.
The purple background highlights the best MSPCG result for a given
	number of nodes.  (This table is reproduced from the Ph.D. thesis
	of Jiqun Tu, Columbia University, 2019.)
}
\label{tab:ss}
\end{table}

\section{Ongoing Evolution of the Ensemble}

As of late December, 2019, we have produced 730 unit length
trajectories for the $96^3 \times 192 \times 12$ ensemble on Summit.
With Chulwoo Jung integrating the QUDA enabled versions of the MSPCG and EOFA algorithm
into the Columbia Physics System and managing the production
running, he was able to refine the choices of
Hasenbusch masses during the first few hundred
trajectories.  Even after around 200 trajectories,
when the lattice should have been close to thermalized,
he noticed that $\langle \exp(-\Delta H) \rangle$ was
not averaging to one, with occasional large values, ${\cal O}(20)$,
appearing.  We adjusted the number of poles in the rational
approximation needed in the EOFA heat bath, and tightened the
stopping condition for the sovers calculating this part of the
Hamiltonian, and the occasional outliers have gone away.  In
particular for trajectories 300 to 730,
$\langle \exp(-\Delta H) \rangle = 0.94(5)$, where the
error comes from assuming that values of $\Delta H$ are decorrelated
for each trajectory.

An important quantity is the evolution of global topological charge, $Q_{top}$,
which is shown in Figure 2.  From this figure, one can see
substantial excursions in $Q_{top}$ that take ${\cal O}(200)$ molecular
dynamics time units.  The histogram of topology shows it is not symmetric
around zero, with a current bias towards positive values.  This figure
indicates that we are far from a situation with frozen $Q_{top}$ and we
anticipate running with longer molecular dynamics trajectories to further
speed up the evolutions of topological charge.

\begin{figure}
\begin{center}
\includegraphics[scale=0.55]{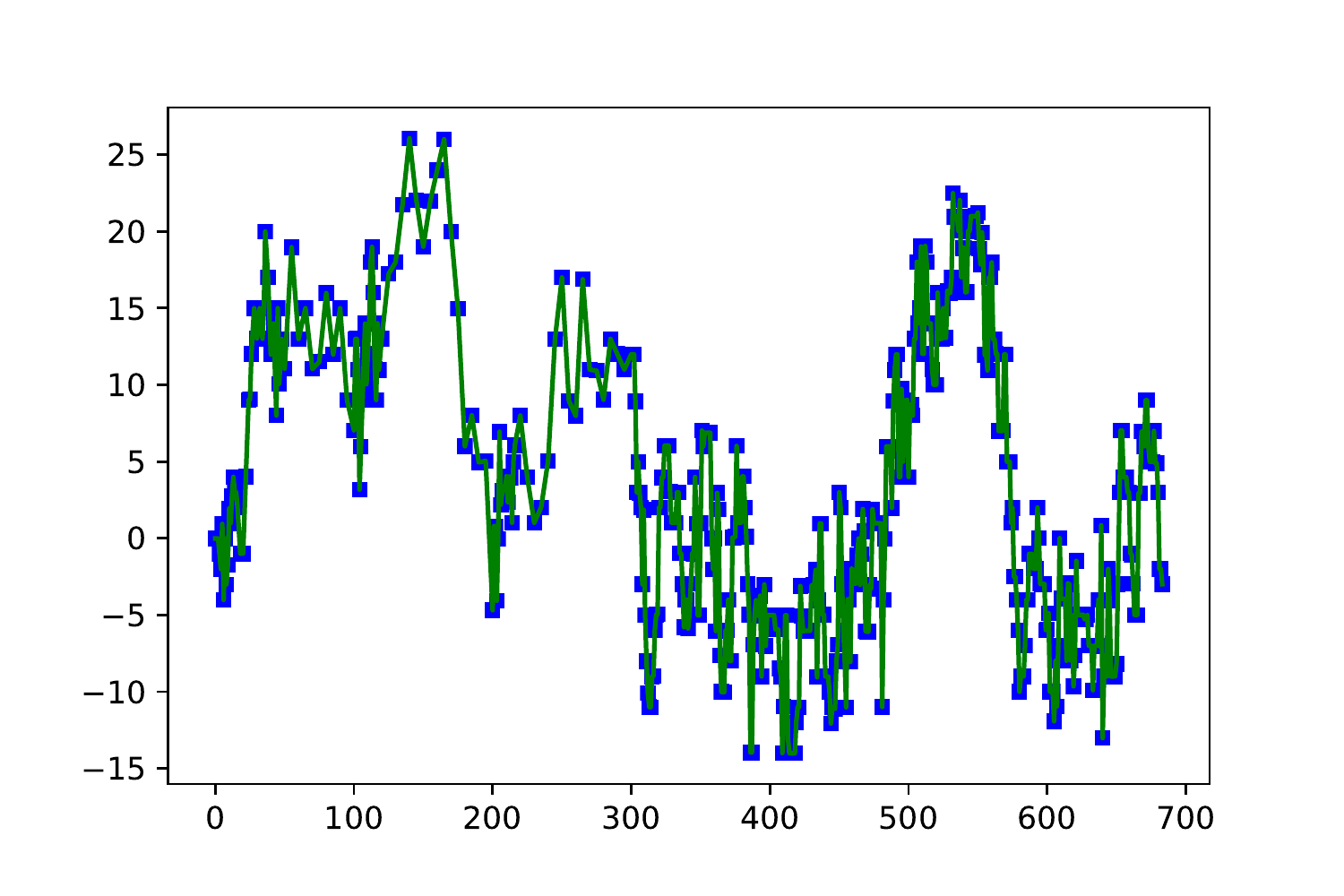}
	\caption{The evolution of topological charge for the $96^3 \times 192 \times 12$ ensemble.
	The horizontal axis is in molecular dynamics time units and for us one trajectory is one
	molecular dynamics time unit.}
\end{center}
\end{figure}

\section{Conclusions}

We have a successfully evolved a $96^3 \times 192 \times 12$, 2+1 flavor, (M)DWF ensemble
for 730 molecular dynamics time units on the Summit computer at ORNL.  This ensemble
has $1/a = 2.774$ GeV and a spatial volume of (6.9 fm)$^3$.  The evolution has being done
on partition sizes of 256, 512 and 1024 nodes.  We see reasonable evolution of topological charge and
are expecting to move to longer molecular dynamics trajectory lengths to improve
topological charge evolutions.


\begin{thebibliography}{99}
\bibitem{DWF-PHYS} T. Blum, {\em et.\ al.}, Phys.\ Rev.\ D93 (2016) 074505, arXiv:1411.7017 [hep-lat].
\bibitem{DWF2774} Peter A. Boyle, {\em et.\ al.}, JHEP 1712 (2017) 008, arXiv:1701.02644 [hep-lat].
\bibitem{EOFA1}Y.-C. Chen and T.-W. Chiu (TWQCD), Phys.\ Lett.\ B738, 55 (2014), arXiv:1403.1683 [hep-lat].
\bibitem{EOFA-RBC} C. Jung, C. Kelly, R.D. Mawhinney and D.J. Murphy, Phys.\ Rev.\ D97 (2018) 054503,
	arXiv:1706.05843 [hep-lat].
\bibitem{MSPCG} D. Guo, R. Mawhinney and J. Tu, arXiv:1804.08593 [hep-lat].
\bibitem{MSPCG-JT} Jiqun Tu, arXiv:1811.08488 [hep-lat], PoS LATTICE2018 (2018) 030.
\bibitem{MS} D.P. O'Leary and R.E. White, SIAM J. Algebr. Discret. Methods 6, 630 (1985).

\end{thebibliography}
\end{document}